\documentclass[preprint]{aastex} 

\newcommand{\com}[1]{}
\usepackage{amsmath,amssymb,amsfonts}
\usepackage{graphicx}
\usepackage[update,prepend]{epstopdf}
\usepackage{hyperref}
\usepackage{array}
\usepackage{cite}
\usepackage{natbib}
\usepackage{morefloats}
\usepackage{color,soul}

\begin{document}

\title{Simultaneous Observations of Giant Pulses from Pulsar PSR~B0031-07 at 38~MHz and 74~MHz}

\author{
Jr-Wei Tsai\altaffilmark{1},
John H. Simonetti\altaffilmark{1},
Brandon Bear\altaffilmark{1},
Jonathan D. Gough\altaffilmark{2},
Joseph R. Newton\altaffilmark{3},
Michael Kavic\altaffilmark{4}
}

\altaffiltext{1}{Department of Physics, Virginia Tech, Blacksburg, VA 24061, U.S.A}
\altaffiltext{2}{Department of Chemistry, Lehman College, CUNY, Bronx, New York 10468, U.S.A.}
\altaffiltext{3}{Department of Chemistry and Physics, Georgia Regents University, Augusta, GA 30912, U.S.A.}
\altaffiltext{4}{Department of Physics, Long Island University, Brooklyn, New York 11201, U.S.A.}


\begin{abstract}
The first station of the Long Wavelength Array (LWA1) was used to study PSR~B0031-07 with simultaneous observations at 38 and 74~MHz. We found that 158 (0.35\%) of the observed pulses at 38~MHz and 221 (0.49\%) of the observed pulses at 74~MHz qualified as giant pulses in a total of 12 hours of observations.  Giant pulses are defined as having flux densities of a factor of $\geq$ 90 times that of an average pulse at 38~MHz and $\geq$ 80 times that of an average pulse at 74~MHz.  The cumulative distribution of pulse strength follows a power law, with an index of $-$4.2 at 38~MHz and $-$4.9 at 74~MHz.  This distribution has a much more gradual slope than would be expected if observing the tail of a Gaussian distribution of normal pulses.  The dispersion measure value which resulted in the largest signal-to-noise for dedispersed pulses was DM $=10.9$~pc~cm$^{-3}$. No other transient pulses were detected in the data in the wide dispersion measure range from 1 to 5000~pc~cm$^{-3}$. There were 12 giant pulses detected within the same period from both 38 and 74~MHz, meaning that the majority of them are not generated in a wide band.

\end{abstract}


\pagebreak 

\section{Introduction}
The pulsar emission from PSR~B0031-07 exhibits many unusual properties, such as systematically drifting sub-pulses across the profile of an observed average pulse (AP) (e.g. \citet{1970ApJ...162..727H}, \citet{2011A&A...525A..55K}).  PSR~B0031-07 is an example of a pulsar which has been observed to emit giant pulses (GPs) but which possesses a relatively low magnetic field in the light cylinder (B$_{LC}$) \citep{2004AstL...30..247K}.

The dispersion measure (DM) of PSR~B0031-07 shows an unexpectedly large variance from 10.89 to 11.38~pc~cm$^{-3}$ \citep{2011A&A...525A..55K, 2013MNRAS.431.3624Z}.  \citet{2013MNRAS.431.3624Z} argued that at low observing frequencies the DM can be determined with higher accuracy given that dispersion is proportional to the inverse square of frequency. Observations of pulsar pulse emission can be used as an effective probe of the interstellar medium (ISM). However, the effect of interstellar scattering along the line-of-sight to this pulsar is relatively unexplored. In the current work we analyze simultaneous observations of PSR~B0031-07 at 38 and 74~MHz. Our results include observation of GPs, a determination of the pulsar's DM and a measurement of the scatter time for the ISM along the line-of-sight to the pulsar. These observations were conducted using the first station of the Long Wavelength Array (LWA1) \citep{2013ITAP...61.2540E}.

GPs have been detected from PSR~B0031-07 at 40~MHz \citep{2004A&A...427..575K} and 1.54 ~GHz \citep{2011RAA....11..974T}, with a similar power law index $-4.5$ of the cumulative number distribution at 40~MHz.  However, the GP's intensity ratio with AP's is much weaker at 1.54~GHz than at frequencies below 100~MHz.  Our observations show a similar intensity and power law distribution for GPs when compared to other observations below 100~MHz, which indicate that GPs maintain a similar intensity ratio to APs between 38 and 74~MHz.  We detected 158 (0.35$\%$) GPs at 38~MHz and 221 (0.49$\%$) GPs at 74 MHz. No other transient pulses were detected in a wide dispersion measure range from 1 to 5000 pc~cm$^{-3}$. For the observations detailed here the DM value which resulted in the largest signal-to-noise for dedispersed pulses was chosen; this was DM $=10.9$~pc~cm$^{-3}$.

\section{Observations and Data Reduction}\label{obanddr}
This work utilizes the LWA1 \citep{2013ITAP...61.2540E} telescope array in central New Mexico, which operates in the frequency range 10--88~MHz.  This array is made up of 261 dual-polarized antennas, consisting of 256 antennas spread over an area of 110~m by 100~m and 5 outlier antennas ranging from 200--500~m from the center of the core.  The dipole outputs are digitized individually, and the DRX beam-forming mode is formed by four fully independent beams.  Each beam is independently tunable with a bandwidth of 19.6~MHz and a range of 10--88~MHz.  Pointing at the zenith, the full-width at half-maximum (FWHM) bandwidth at 74~MHz is approximately $4.3^\circ$.  This FWHM depends on the frequency as $\nu^{-1.5}$.  The beam sensitivity of the LWA1 is dependent on the local sidereal time and direction of the beam (and source) because the system temperature is strongly dependent on Galactic emission.
The LWA1 is able to observe the profile evolution of individual pulses between two observing frequencies, as a result of its capacity to observe at two frequencies simultaneously \citep{2013ITAP...61.2540E}.

The LWA1 was used for observations of PSR~B0031-07 over the span of three days, for 4 consecutive hours each day.  These observations were performed from July 26 through July 28, 2013, with each observation beginning 110 minutes before the pulsar passed the meridian.  Two polarizations were recorded as part these observations. These two polarizations were summed before further analysis in order to increase the intensity of observed signals.  A 4096 Fast Fourier Transform was performed on the raw data in steps of 0.209~ms, dividing the 19.6~MHz observing bandwidth into channels of 4.785~kHz, using routines from the LWA Software Library (LSL, \citet{Dowell:2012rt}).  Observations were conducted at two frequencies centered at 38 and 74~MHz, each with a bandwidth of 16~MHz. The reduced bandwidth is due to low system responses.

The following method was used to perform radio frequency interference (RFI) mitigation.  First, average spectra for each 2.09~s interval were obtained (10,000 steps of 0.209~ms).  A 16th order polynomial was fit to each average spectrum.  Since there were approximately 16 ripples in the bandpass, a 16th order polynomial was the lowest order which fit the data without over-suppression of the narrow-band RFI. Each average spectrum was then divided by its corresponding 16th order polynomial fit.  Finally, any frequency bin greater than 3$\sigma$ above the mean in any 2.09~s average spectrum was then marked as RFI contaminated in all of the corresponding 0.209~ms spectra.

There exist variations in the shape of the bandpass as a function of time, which is dominated by the diurnal Galactic background variation.  In order to effectively search for transient pulses, this variation in the bandpass must be accounted for.  After masking the RFI contaminated frequency bins, the variation in the shape of the bandpass was determined and removed as follows.  First a median spectrum was computed for each 2.09~s of data.  Then, 150 of these spectra were combined to compute a median spectrum for approximately 5 minutes of observation.  This was a sufficient time length to smooth out the diurnal Galactic background.  A 101 frequency length boxcar was used to perform a moving boxcar average across each 5 minute spectrum.  Finally, each 0.209~ms spectrum was divided by the boxcar smoothed spectrum corresponding to its epoch.  To remove any end effects of the averaging, the first 360 channels and the last 395 channels were removed, leaving a final data set with a bandwidth of 16.0~MHz.  From this data, spectrograms of frequency (vertical axis) and time (horizontal axis) were produced. 

In searching for pulsar GPs in our data, we use a technique described by \citet{2003ApJ...596.1142C} to search our intensity versus time and frequency data for individual pulses that could be of astrophysical origin.  This method consists of creating dedispersed time series for a range of candidate DMs and smoothing each data set with several increasing averaging-boxcars in order to search for pulses of temporal width corresponding to smoothing time, thus giving the best signal-to-noise ratio for a pulse of interest.  Pulses that are larger in strength and number than what is expected for Gaussian noise in the data are considered pulses of possible astrophysical origin. Though it is possible that these pulses are RFI or other transient non-astrophysical events, it is more probable that candidate pulses of the proper DM for PSR~B0031-07 are pulsar pulses.  Spectrograms were then incoherently dedispersed (summing intensities) and searched over various time series for a total of 28,454 candidate DMs in this manner.  These data sets ranged from 1 to 5000 pc cm$^{-3}$. The DM spacing is given by 
\begin{equation}
\delta \mbox{DM} = \mbox{DM} \frac{\Delta\nu}{B},
\label{eqn:DMspacing}
\end{equation}
where $B$ and $\Delta\nu$ are the bandwidth and channel-width respectively. This results in a temporal smearing across the frequency channel which is equal to the temporal smearing across the DM spacing and cannot be removed.  Smoothing of this time series was then performed in steps by a moving boxcar average with a boxcar length equal to the two time samples.  One of the resulting time samples was then removed.  Increasing smoothness is then efficiently obtained by repeating this smoothing and decimating process, while the resulting time series is searched for pulses at each step.  For each dedispersed time series, this step was performed 15 times and the resulting final time sample duration period for the last-smoothed time series was $2^{15}\times0.2089~{\rm ms} = 6.85$~s.

The only DM for which the resulting time series produced transient events with S/N $\geq$ 5.5 was for a DM of 10.9~pc~cm$^{-3}$, which is the known DM for PSR B0031-07 and thus indicate that these events are indeed a result of a signal from the pulsar.  The number of transient events that we found with S/N $<5.5$ are consistent with what would be expected from Gaussian noise. The number of events of S/N $>5.5$, however, are both greater than expected from Gaussian noise alone, and appear at the DM of the pulsar.  We are therefore confident that using this method of analysis and focusing on transient events with S/N $>5.5$ results in the selection of pulses produced by PSR~B0031-07.  

It is important to recognize that the quoted pulsar flux densities for APs are averages in time which include both energy received during the pulse and between pulses.  When using the Cordes-McLaughlin method, however, the S/N is computed in the time series which is smoothed to the temporal width of the pulse.  This is an appropriate method to use for describing the S/N of GPs, since GPs are isolated in time.  Therefore, when we measure the flux density of our GPs and compare them to APs, we will adopt the more conventional time-average throughout a pulse period.

Finally, as an additional test of the parameters discussed here, we used an incoherent dedispersion routine to vary the DM in a range from 10.845 to 10.945~pc~cm$^{-3}$ while searching one bandpass-corrected 0.209~ms spectra for large individual pulses from PSR B0031-07.  The results of this search showed that a DM = 10.9 pc cm$^{-3}$ produced the highest S/N.  Thus, this DM was chosen for the search for GPs for the remainder of the data set.

\section{Behavior of Pulses}\label{behaviorofpsr}
APs and some GPs exhibited multiple-component profiles and were fit with multiple Gaussian functions.  This allowed us to study both their flux density and phase.  Gaussians were fit at both frequencies, therefore it enabled a simple comparison of the pulse.  Because the observations at these two frequencies were simultaneous, the relative phase could be determined after dedispersion.  The two frequencies were dedispersed independently using the DM of the GP with the largest S/N.

Using the fitted Gaussians, the flux density and phases of components were determined. The flux density is calculated by converting the area under the fitted curves to the system equivalent flux density (SEFD). The flux density ratio of a GP and the AP is the ratio of the area under the fitted curves. The area of the AP was divided by the number of folded periods.

\subsection{Flux density}
Rough flux densities for the AP and GPs were obtained by using an estimated SEFD; drift scans on other objects were not used for calibration. For an observation of 1 Hz bandwidth and an integration time of 1 second, the SEFD is the flux density a source in the beam needs to produce a S/N of unity. Galactic noise is the dominant contribution to system noise at low frequencies.  The model established by Ellingson for estimating the SEFD, which takes account of the combined effects of all sources of noise \citep{Ellingsonsen}, uses a spatially uniform sky brightness temperature $T_B$, dependent on observing frequency $\nu$, where
\begin{equation}
T_b = 9751 \mbox{K} \left(\frac{\nu}{38\mbox{MHz}}\right)^{-2.55} 
\label{eqn:brightness temperature}
\end{equation}
and ignores the ground temperature contribution as negligible.  The receiver noise is about 250~K, but has little influence on the SEFD.  When applied to LWA1, this model shows that the correlation of Galactic noise between antennas significantly desensitizes the array for beam pointings that are not close to the zenith.  Notably it shows that considerable improvement is possible using beam-forming coefficients that are designed to optimize S/N under these conditions. Using strong flux density calibrators,  \citet{2013ITAP...61.2540E} checked this model and found the results roughly correct.  Given that our observations at transit are at non-zero zenith angle and using Ellingson's model and drift scan results we can estimate an appropriate SEFD for use in our observations; accordingly, we estimate an uncertainty of 50\%.  The S/N of pulses away from the moment of transit are corrected by a factor which compensates for decreasing effective collecting area and increasing SEFD, with increasing zenith angle.

Therefore, we assign the flux density to a pulse, as averaged across the entire pulse period, as
\begin{equation}
S = \frac{\rm SEFD}{\sqrt{2B\Delta t}}\ \frac{1}{N_{\rm bins}} \sum_{i=1}^{N_{\rm bins}} \frac{I_i}{\rm rms} = \frac{\rm SEFD}{\sqrt{2B\Delta t}}\ {\rm \overline{S/N}}
\label{eqn:flux density}
\end{equation}
where $B$ in~Hz is the bandwidth, $\Delta t$ in seconds is the duration of a time sample, $N_{\rm bins}$ is the number of time samples (bins) in a pulse period, the sum is over the full pulse period, the $I_i$ are the intensity values (arbitrary units) in the Gaussian pulse profile fitted to a pulse (a baseline average was already subtracted from the data), rms is measured in the baseline, and the $\rm \overline{S/N}$ is the average S/N during the pulse period.  Using \citet{2013ITAP...61.2540E} we assume the same SEFD is 15,000~Jy ($\pm$50\%) for both observing frequencies while transiting at a zenith angle of about 26$^\circ$.

\begin{figure}
\begin{center}
\includegraphics[width=.45\textwidth]{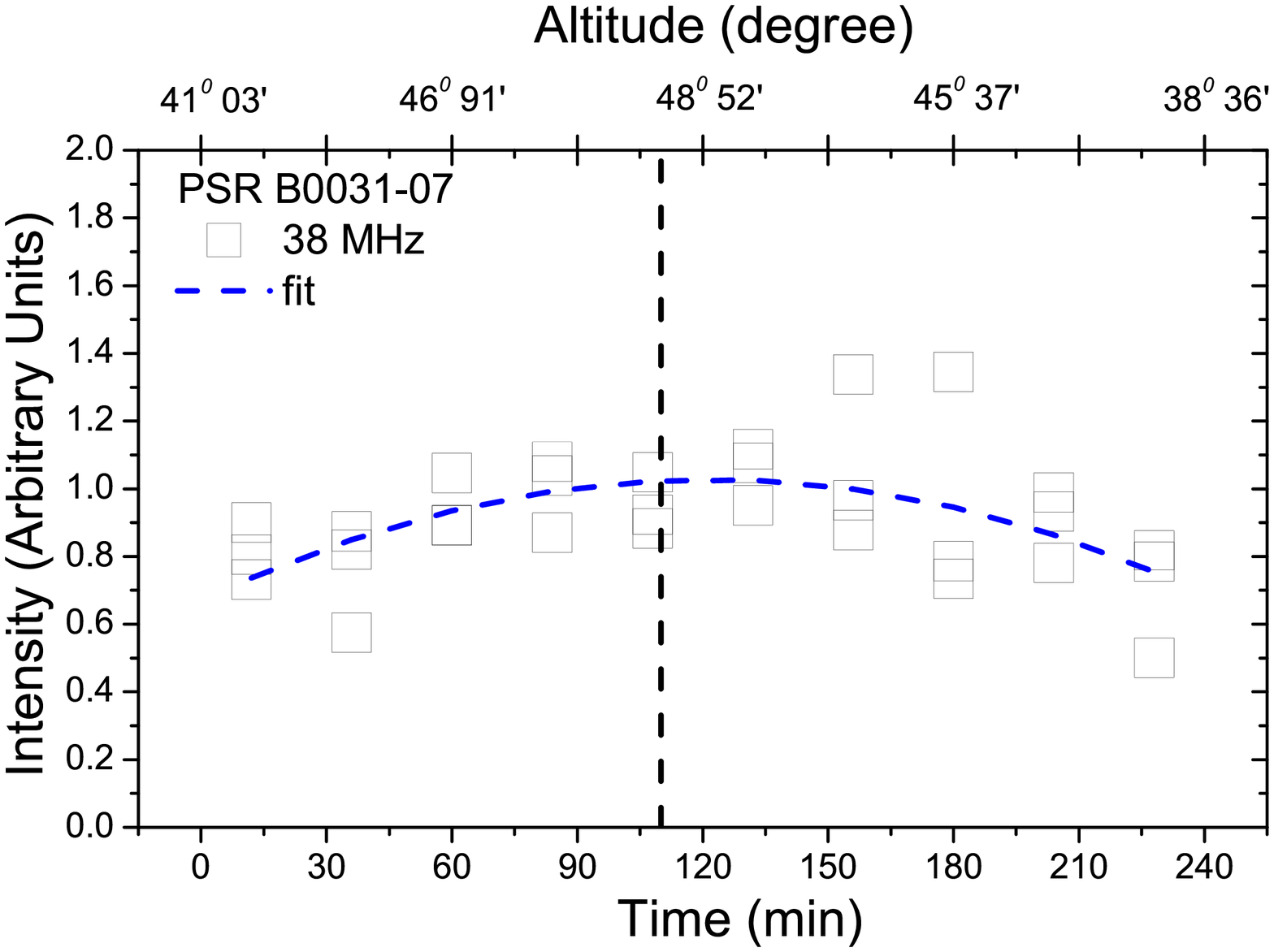}
\includegraphics[width=.45\textwidth]{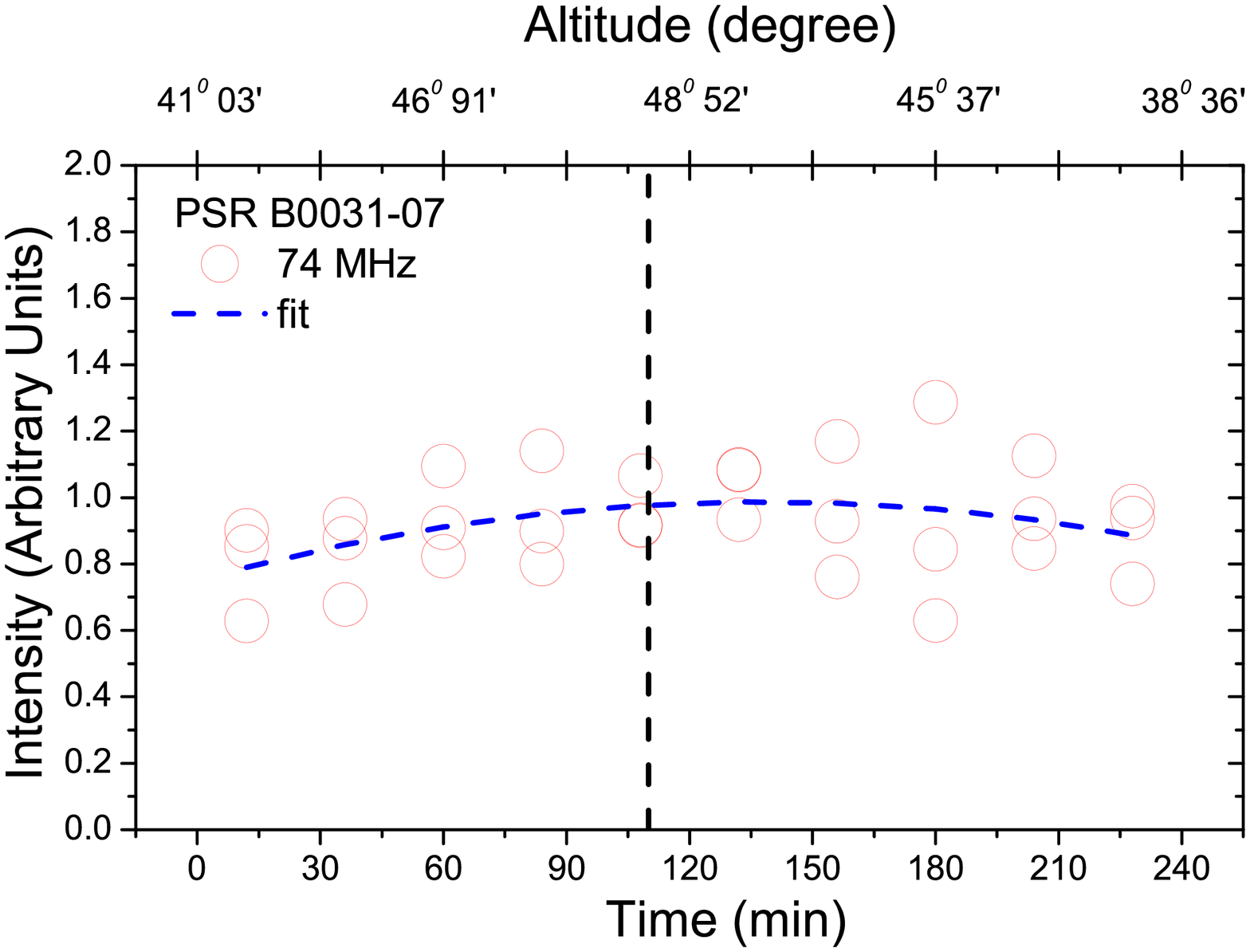}\\
\caption{The flux density of the LWA1 measurement as a function of time for observation frequencies of 38~MHz (left) and 74~MHz (right), collected for a duration of 4 hours on three consecutive days in beam tracking mode.  The observations began each day 110 minutes before PSR B0031-07 passed the meridian at an elevation of 48.58$^\circ$.  Each individual data point represents an average intensity over a time duration of 24 minutes and there are three data points plotted for each observation time corresponding to three consecutive days of observation.  The intensity signals show a clear dependence on the target's zenith angle for both observation frequencies.  These systematic effects were accounted for when calculating the intensity of GPs.
}
\label{altitudecorr}
\end{center}
\end{figure}

It should be noted that the sensitivity of LWA1 is dependent on the target elevation.  This effect was corrected by calculating sequences of 24~minute averages of the pulse flux densities and fitting the variation with a polynomial function, with a maximum value of unity at or near the meridian, as shown in Figure \ref{altitudecorr}.  As such, the flux of any particular pulse was divided by the polynomial to remove the zenith angle dependence for subsequent analysis.

\subsection{Profile of PSR~B0031-07}\label{profile0031}
\begin{figure}
\begin{center}
\includegraphics[width=.75\textwidth]{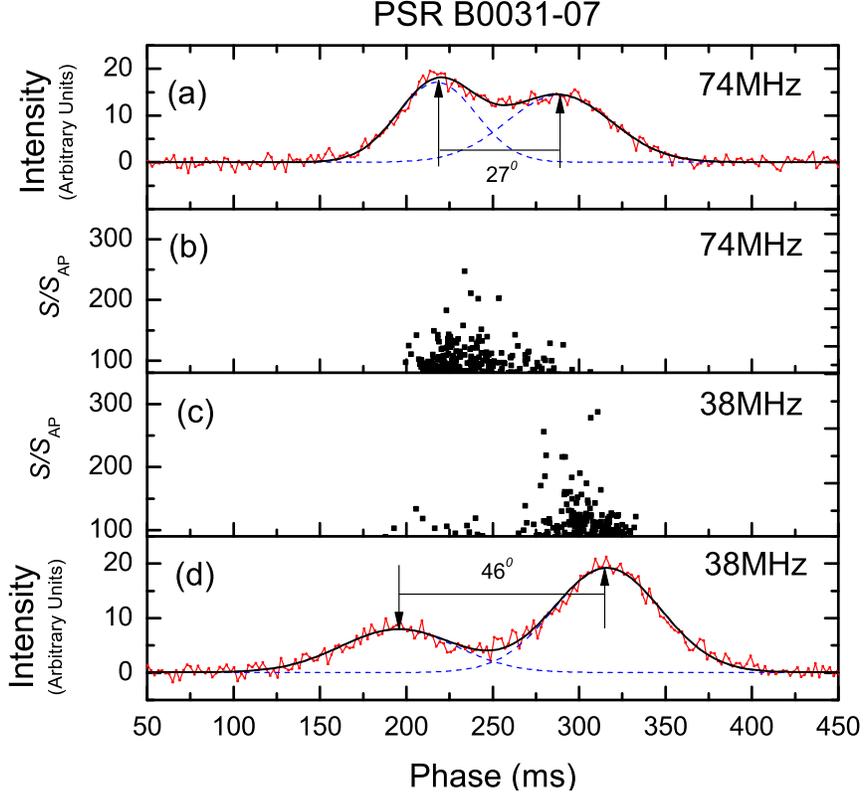}
\caption{Plot of the AP profiles and GP peak phases for frequencies of 74~MHz and 38 MHz.  Panels (a) and (d) show the profile of the AP at 74~MHz and 38~MHz respectively, with dashed lines to represent the two individual Gaussian fits to the profile.  The black solid curve is then the sum of the two Gaussian curves.  Panels (b) and (c) show GP peak phases and flux density at frequencies of 74 MHz and 38 MHz respectively compared to the flux density of the AP.  A GP is defined as a pulse with flux density ratio ($S/S_{\rm AP})>90$ at 38 MHz and  ($S/S_{\rm AP})>80$ at 74~MHz.  The component peak intensity ratio is 1.25 and 0.41 at 74 and 38~MHz respectively. The pulsar's period is 942.87~ms.
}
\label{phasesingledouble}
\end{center}
\end{figure}

\begin{figure}
\begin{center}
\includegraphics[width=.75\textwidth]{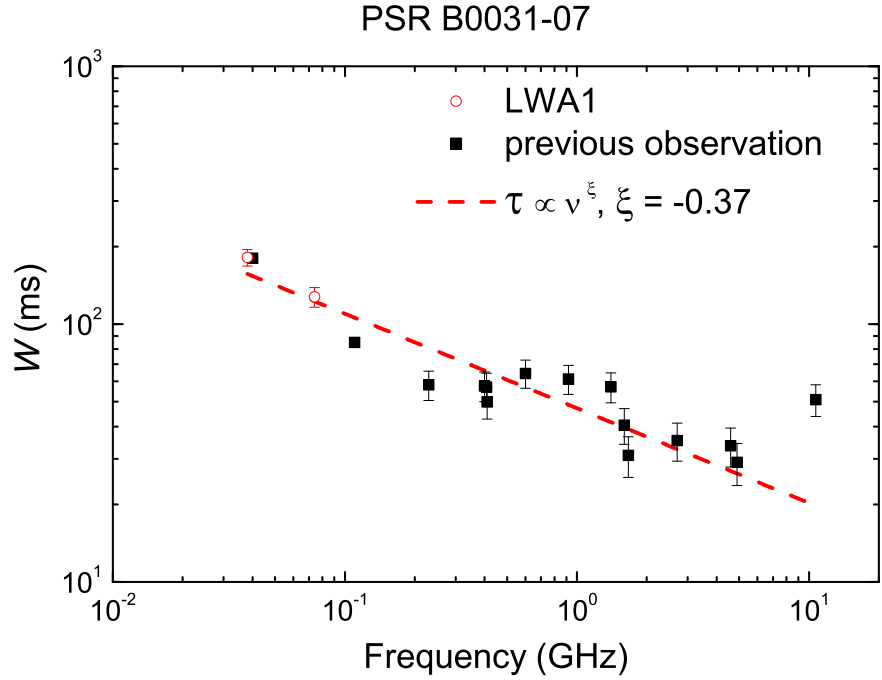}
\caption{The collection of pulse widths observed for PSR~B0031-07 includes both data included in this work and data from the literature.  The pulsar's period is 942.87~ms.  The error bar is specified along with the references for the observations cited.  The red dashed line is fitted from 38~MHz.  Observations are from \citet{1971ApJS...23..283M}:0.41 and 1.665GHz. \citet{1975A&A....38..169S}:2.7 and 4.9~GHz. \citet{1986A&A...161..183K}: 4.6 and 10.7~GHz. \citet{2004A&A...427..575K}: 40 and 111~MHz. and this work: 38 and 74~MHz.
}
\label{scattering}
\end{center}
\end{figure}

As noted above, two Gaussian functions were required for an adequate fit to the pulse profiles observed (Figure \ref{phasesingledouble}).  Comparing the AP profile at the two observing frequencies reveals that both components shift away from center.  Previous observations of PSR~B0031-07 below 100~MHz showed strong profile evolution for both components.  To compare with the previous observations by \citet{2004A&A...427..575K} at 40~MHz, we adopt the same definition of the width of the AP at 38 and 74~MHz. Thus we define the pulse width as the FWHM across both components \citep{2004A&A...427..575K}.  The  profile width ($W$) of the AP is 181.4~ms and 127.5~ms at 38 and 74~MHz respectively.  Figure \ref{scattering} shows $W$ for frequencies above 40~MHz. In the context of other observations of PSR~B0031-07 at frequencies from 38~MHz to 10.7~GHz, we observed a temporal broadening spectral index of $\xi=-0.37\pm0.07$ (Figure \ref{scattering}).

Our profile width measurement at 38~MHz is comparable to \citet{2004A&A...427..575K} at 40~MHz (see Figure \ref{scattering}).  Our observations indicate pulses from PSR~0031-07 are broadened at low frequencies. This is a common for many pulsar observations (e.g. \citealt{1986ApJ...311..684H} and related papers), which can be used to study the beam structures (e.g. radius-to-frequency mapping \citealt{1978ApJ...222.1006C}).  The temporal broadening spectral index of PSR~B0031-07 is comparable to many pulsars (e.g. Figure 3 of \citealt{2015arXiv150906396P} for comparison). One could estimate the emission altitude at 38 and 74~MHz from aberration and retardation effects (see the discussion in \citealt{2012A&A...543A..66H}).  However, the impact angle and inclination angle between the magnetic and rotational axis of PSR~B0031-07 have not been determined, Thus we are not able estimate the emission altitude at 38 and 74~MHz in this study.

\begin{figure}
\begin{center}
\includegraphics[width=.75\textwidth]{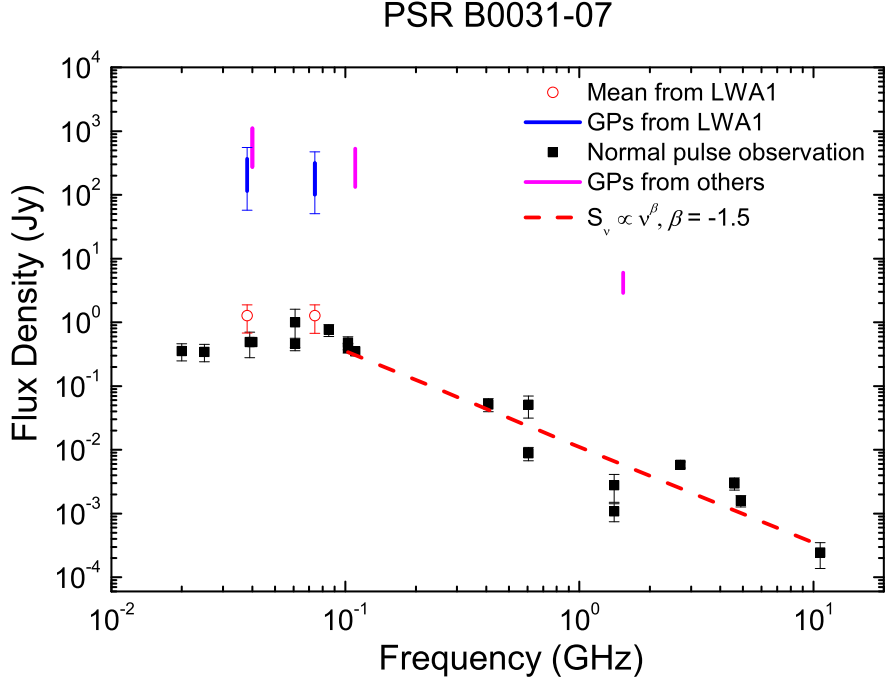}
\caption{A collection of observed flux densities from the literature and current data from this work for PSR~B0031-07 at different observing frequencies are shown here.  Unless otherwise specified in the cited reference, errors of 50\% have been assumed for all AP flux density points.  The black solid squares represent the flux density for the AP.  The red halo circles represent the values for LWA1 at 38 and 74~MHz (this work). For the GP data, the thick vertical line indicates the range of observed values and the thin error bars extending above and below indicate the 50\% error associated with the lowest GP and highest GP flux density.  The dashed red line is a linear fit to the log-log data points for AP observations above 0.1~GHz.  Observations include
\citet{2011ChA&A..35...37N}:1.54~GHz, 
\citet{1971ApJS...23..283M}:0.41 and 1.665~GHz,
\citet{1975A&A....38..169S}:2.7 and 4.9~GHz, 
\citet{1979SvA....23..179I}: 61 and 102.5~MHz, 
\citet{2004A&A...427..575K}:40 and 111~MHz, 
\citet{1981AJ.....86..418R}: 430~MHz, 
\citet{1986A&A...161..183K}: 4.6 and 10.7~GHz, 
\citet{1992ApJ...385..273P}: 25, 47, 112, 430, 1408 and 4800~MHz, 
\citet{1995A&A...301..182R}: 45~MHz, 
\citet{2013MNRAS.431.3624Z}:20 and 25~MHz,
\citet{1995MNRAS.273..411L}: 606 and 1408~MHz
and this work: 38 and 74~MHz.
Note that the GP flux densities from (\citet{2011ChA&A..35...37N} and \citet{2004A&A...427..575K}) at 40, 111 and 1540~MHz are peak averages. In order to compare to their results, we multiplied our period averaged result with the ratio of $\frac{period}{W}$ to obtain the peak averaged flux density.
} 
\label{fluxspectra}
\end{center}
\end{figure}

The period-averaged flux densities are 1.28 Jy and 1.27 Jy at 38 and 74~MHz, respectively. The flux densities for APs at frequencies of 20~MHz and above are shown in Figure \ref{fluxspectra}, along with observations of GPs by \citet{2004A&A...427..575K} at 40 and 111~MHz and \citet{2011ChA&A..35...37N} at 1.54~GHz.  For comparison, we have added data points from this work.  The error bar on our AP flux densities indicates our $\sim$50\% uncertainty. The value of spectral index $\beta$, where S$_\nu \propto \nu^\beta$, determined for frequencies above 0.1~GHz is $-1.5\pm0.2$. This values is similar to the value of $-1.4\pm0.11$ found by \citet{2000A&AS..147..195M} for frequencies above 0.4~GHz. Spectral turnover can be seen around 100~MHz as reported by \citet{1973A&A....28..237S}. This is consistent with our observations and can bee seen in figure \ref{fluxspectra}.

\subsection{GPs from PSR~B0031-07}\label{gps0031}
\begin{figure}
\begin{center}
\includegraphics[width=.75\textwidth]{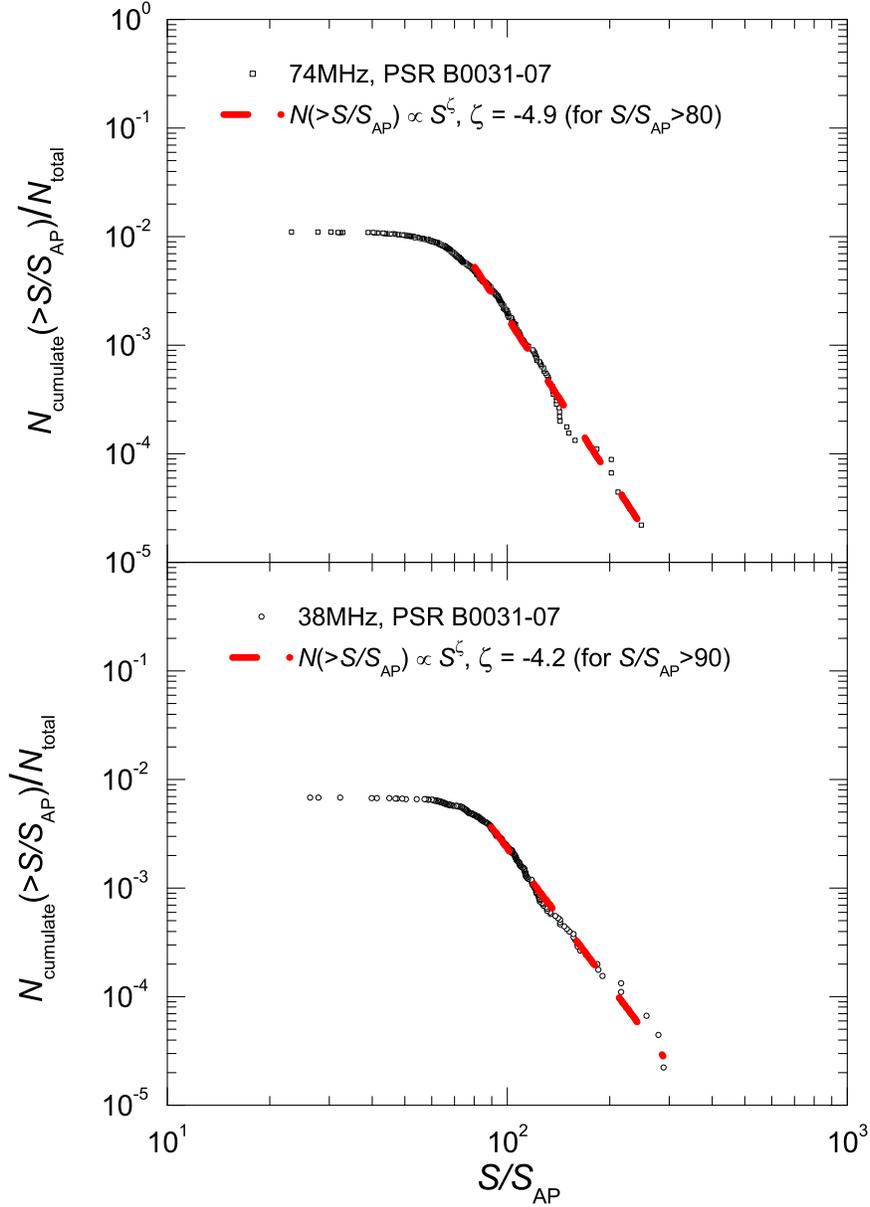}
\caption{(Top) The pulse cumulative number $N(>S/S_{\rm AP}$) for observation frequency 74~MHz as a function of flux density is plotted as black squares, with the power-law fit for $S/S_{\rm AP}>80$ included as a red curve. (Bottom) The pulse cumulative number $N(>S/S_{\rm AP}$) for observation frequency 38~MHz as a function of flux density is plotted as black circles, with the power-law fit for $S/S_{\rm AP}>90$ included as a red curve.
}
\label{cumuindex}
\end{center}
\end{figure}

We adopted a definition for GPs from PSR~B0031-07 similar to the one found in \citet{2004A&A...427..575K}.   They found the minimum flux density ratio as 100 at 40~MHz with a cumulative distribution index of $-4.5$.  This power law distribution yields a flux density ratio of $\gtrsim$ 90 at 38~MHz and $\gtrsim$ 80 at 74~MHz for our observations, as shown in Figure \ref{cumuindex}.  Therefore, a GP is defined by the two different flux density ratio ranges over which the two power laws apply.  The index was found to be $= -4.2$ at 38~MHz and $-4.9$ at 74~MHz, respectively.  The cumulative distribution of the ratio of the observed GP peak flux density to the AP peak flux density for all observed pulses is shown in Figure \ref{cumuindex}.  Our results combined with \citet{2004A&A...427..575K} show that the GP cumulative distribution follows a power law with a similar minimum flux density ratio of $90 \pm 10$ below 80~MHz. This is significantly different than the the minimum flux density ratio of 30 found at 111~MHz \citet{2004A&A...427..575K}.

\begin{figure}
\begin{center}
\includegraphics[width=.75\textwidth]{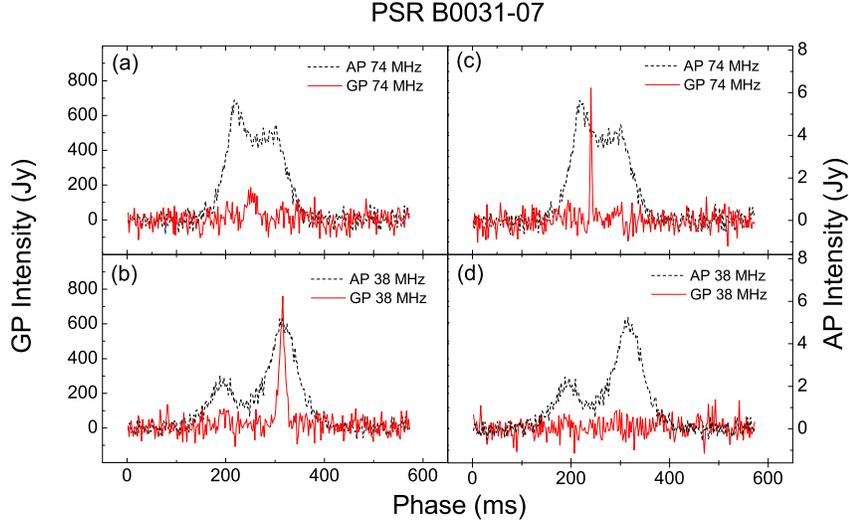}
\caption{
The phase of the GP together with the AP.  Plots of GPs and APs are presented on different scales. Panels (a) and (d) shows the time series during the sameperiod for the second observing frequency.
}
\label{GPs}
\end{center}
\end{figure}

The strongest GP at 38~MHz had a flux density 287 times larger than the AP's, while at 74~MHz the strongest one was found to be 247 times larger.  Figure \ref{GPs} shows examples of two GPs at 38 and 74~MHz which have the largest peak S/N in their dedispersed time series for the observations presented here.  We detected only 2 and 19 double peaked GPs at 38 and 74~MHz respectively, while the majority of GPs have a single peak.  Only 12 GPs were detected in both frequencies within in same periods and not all of them have the same phase. The rest of GPs can only be detected in the one frequency.  The AP profiles and GP peak phases are shown in Figure \ref{phasesingledouble}. The average FWHM of GPs is 10.7~ms at 38~MHz and 5.4~ms at 74~MHz.  While the FWHM of the AP is 181.4~ms at 38~MHz and 127.5~ms at 74~MHz, the average FWHM of GPs are only 1/17 and 1/24 of the AP at 38 and 74~MHz respectively.

\begin{figure}
\begin{center}
\includegraphics[width=.45\textwidth]{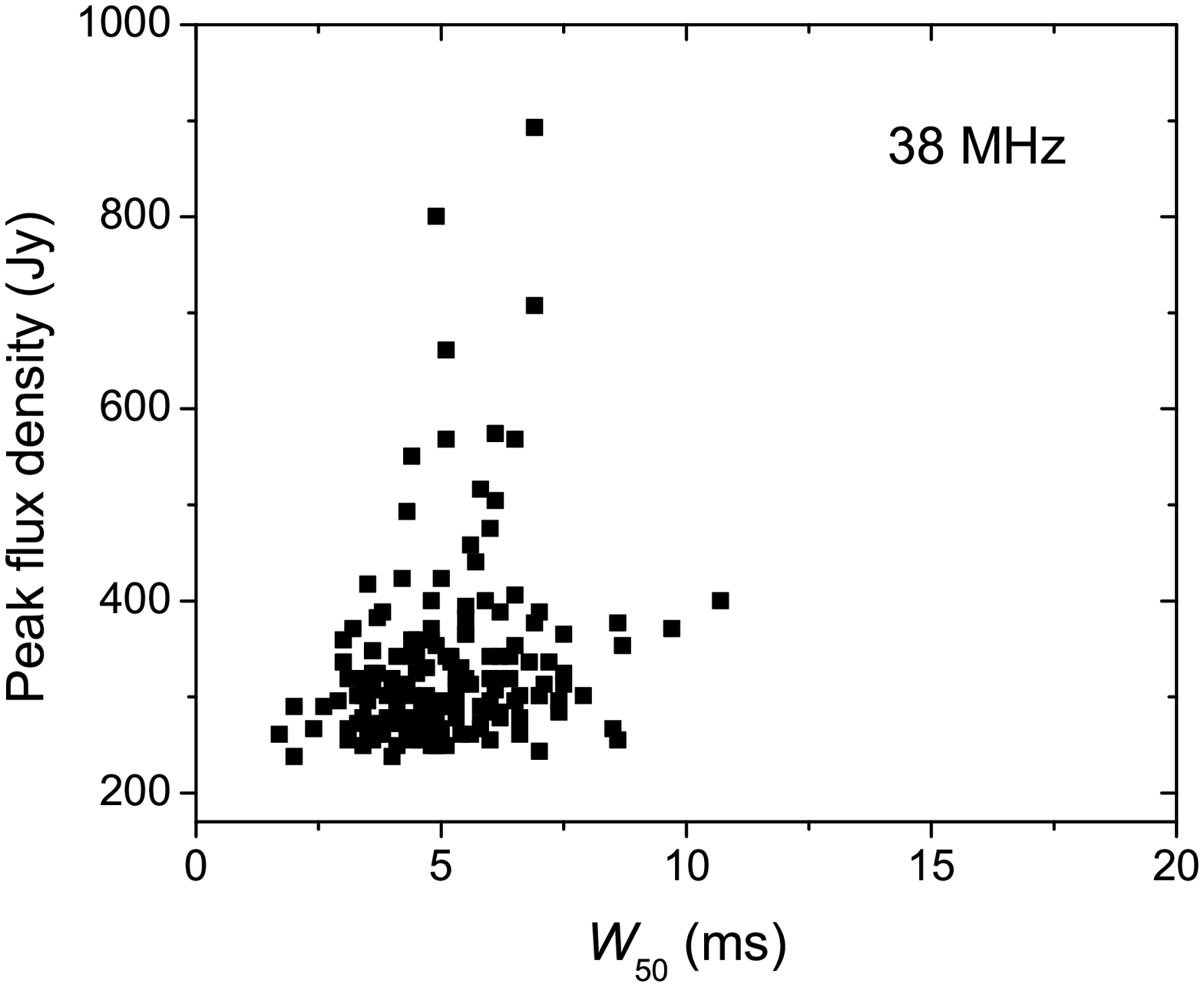}
\includegraphics[width=.45\textwidth]{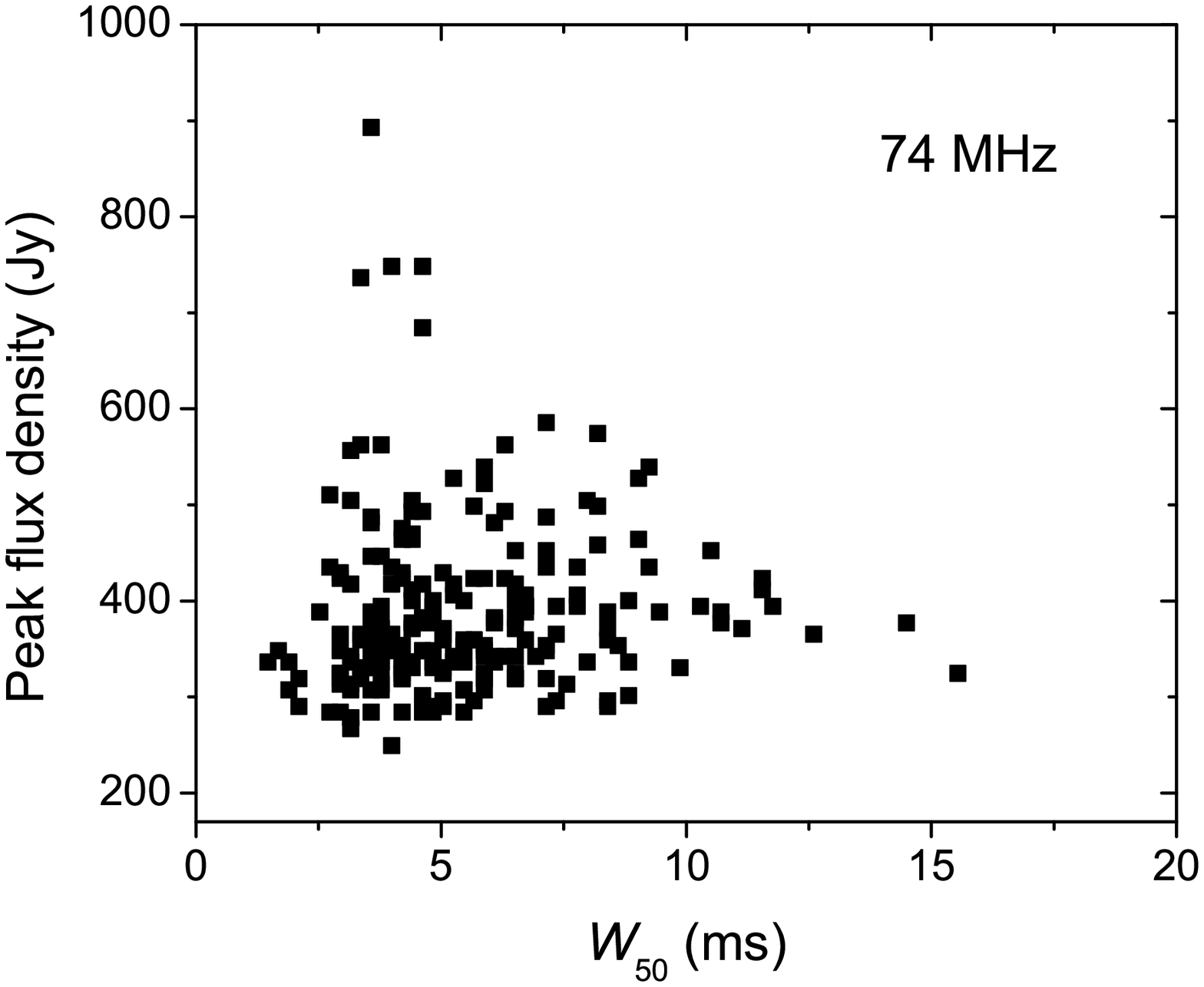}\\
\caption{
The peak flux density of giant pulses versus FWHM ($W_{50}$) at 38 and 74~MHz.
}
\label{snrwidth}
\end{center}
\end{figure}

The statistical properties of GPs from the Crab pulsar found by \citet{2007A&A...470.1003P} indicate that the strongest GPs have a shorter duration than other GPs.  Our observations of GPs from PSR~B0031-07 show that a GP with stronger peak flux density could have shorter duration at 74~MHz, but this trend can not be seen at 38~MHz (see Figure \ref{snrwidth}).

\begin{figure}
\begin{center}
\includegraphics[width=.75\textwidth]{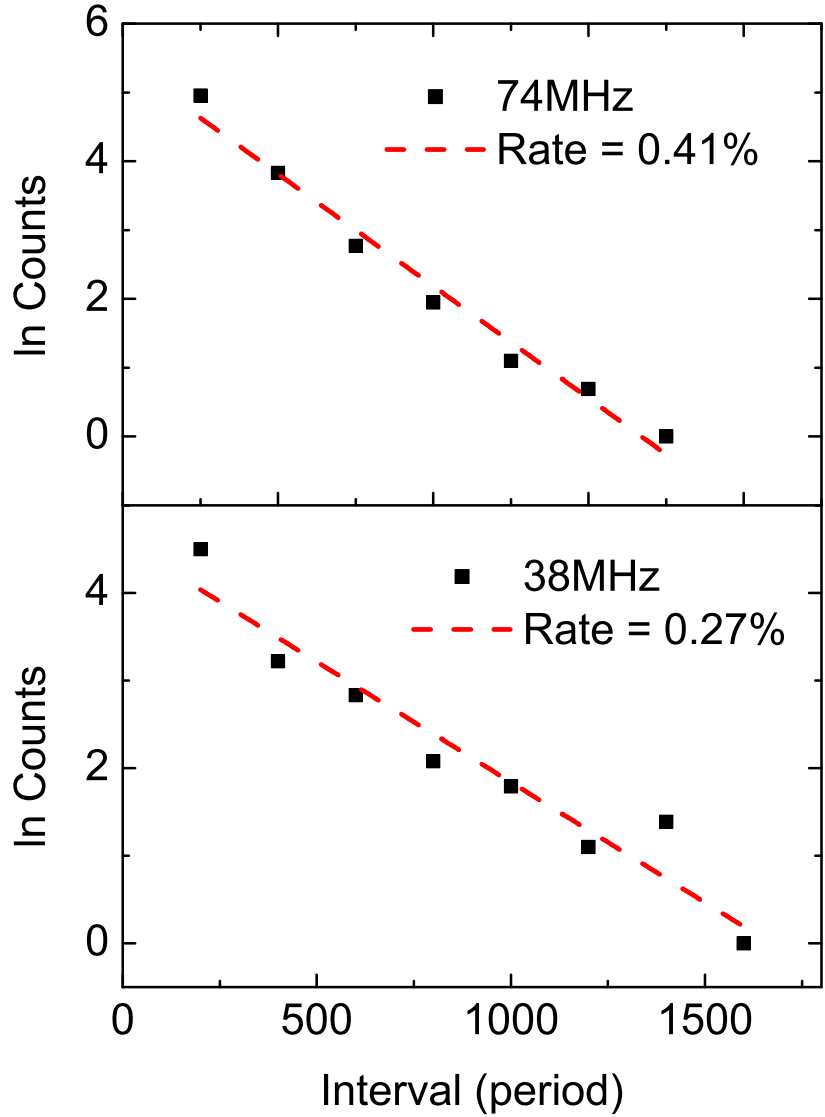}
\caption{The top panel is the plot of the GP interval at a frequency of 74~MHz with a horizontal scale consisting of individual units of ten periods.  The rate is the slope of the dashed line fit from 200 to 1400 periods.  The bottom panel is the plot of the GP interval at a frequency of 38~MHz with the same ten period units for the horizontal scale.  The rate is the slope of the dashed line fit from 200 to 1600 periods.  We did not find interesting clustering or periodic behaviors regardless of the chosen binning interval.
}
\label{intervaldistribution}
\end{center}
\end{figure}

\citet{2012AJ....144..155S} reported that PSR~B0950+08 shows ``quasi-null'' GP emission states in daily 32 minutes observations. GPs from PSR~J1752+2359 show a deviation from Poisson statistics \citep{2006ChJAS...6b..41K}.  We expect the distribution of intervals between GPs to exhibit an exponential functional dependence, i.e. Poisson statistics if each GP is an independent event with a uniform probability for pulse generation at any given time.  The distribution of intervals for one independent event as a function of the number of periods is given by
\begin{equation}
I_1(p) = r\ \mbox{exp}(-r p),
\label{eqn:exponential distribution}
\end{equation}
where $r$ is the average event rate (GPs per period), and the period $p$, for simplicity, is used as a time measure.  Our observations of GPs from PSR~B0031-07 exhibit a Poisson distribution at both 38 and 74~MHz, as shown in Figure \ref{intervaldistribution}.

\subsection{Scattering Analysis}\label{sect:scattering}
Pulsar observations allow for study of the ISM.  We attempted to determine the scatter time $\tau_d$ for both APs and individual pulses with signal-to-noise ratio (S/N) $>5.5$ using the CLEAN-based algorithm \citep{2003ApJ...584..782B}. 
The observed pulses are composed of the intrinsic pulse convolved with propagation effects and instrumental response.  Because the CLEAN-based algorithm utilizes an accumulated delta-like signal to restore the intrinsic pulse, it allows for the deconvolution of various profile shapes without knowledge of the intrinsic profile.  The recorded signal $P_{obs}$ is
\begin{equation} 
P_{obs}(t) = I(t)\otimes \rm{PBF}(t)\otimes r(t),
\label{convolutions}
\end{equation}
where $\otimes$ denotes the convolution, $I(t)$ is the intrinsic pulse profile, $\rm{PBF}(t)$ is the pulse-broadening function, and $r(t)$ is a function which gives the combined instrumental responses including effects due to data reduction.  We include as part of the instrumental responses the effect derived from setting the bin size of the time series to be the same as the temporal resolution and a similar effect for the frequency channel width.  For simplicity, we compare only the exponential decay time $\tau_d$ from a thin screen model, $\rm{PBF}(t)=exp(\frac{-t}{\tau_d})$, for APs with results for pulses with S/N $>5.5$.

We found the exponential decay time from individual pulses with S/N $>5.5$ is 6.1$\pm$2.2~ms and 3.9$\pm$1.4~ms for 38 and 74~MHz, respectively. For the APs we found 39$\pm$7.6~ms and 36$\pm$8.8~ms for 38 and 74~MHz, respectively.  For APs, the scatter time frequency scaling index was found to be $\alpha=-0.18\pm$0.06 and for individual pulses with S/N $>5.5$, the scatter time frequency scaling index is $-1.5\pm$0.2, where $\tau_d\propto\nu^\alpha$.  However, these measured values of $\alpha$ most likely do not reflect the actual scattering index as scatter-broadening should be a property of the ISM through which the pulse is traveling, and therefore should be the same for all types of pulses from a given pulsar.  The inconsistency between the values for $\tau_d$ for APs and for those with S/N$>5.5$ is most likely a result of evolution of the profiles at different frequencies, such as radius-to-frequency mapping \citep{1978ApJ...222.1006C, 2003A&A...397..969K}, rather than the scattering effect from ISM.  Also as \citet{2013MNRAS.434...69L,2015ApJ...804...23K} argued one can obtain reliable values of $\tau_d$ only when the scatter time is significantly larger than the width of the profile. The exponential decay time derived from CLEAN method is much smaller than the profile width. Thus the ISM scattering effect is not strong enough to be measured at the frequencies that we observed along the line-of-sight to PSR~B0031-07.

Furthermore, several studies of the scatter time frequency scaling index \citep{2001ApJ...562L.157L,2004ASSL..315..327L,2013MNRAS.434...69L,2015MNRAS.449.1570L} revealed that only a handful of pulsars have their scaling indices close to the value of $-3.0$. All of these extremes happen only for high DM pulsars for which the ISM can be expected to have some extreme scattering geometry which causes the scaling index to flatten \citep{1999ApJ...517..299L}. PSR B0031-07 is not a high DM pulsar so this should not apply.

\section{Conclusion}\label{sect:conclusion}

GPs and APs were detected while observing PSR~B0031-07 simultaneously at 38 and 74 MHz using the LWA1. Notably, we have shown that GPs and APs from PSR~B0031-07 are weaker than at $\sim$100~MHz. The rate and strength of giant pulses are comparable with other observations at 111~MHz and 40~MHz by \citet{2004A&A...427..575K}. Most GPs can only be detected at one frequency which implies that most GP emission is not wide band. We found that a DM = 10.9 pc cm$^{-3}$ produced the highest S/N for observed pulses.  The CLEAN-based algorithm \citep{2003ApJ...584..782B} was used to analyze the effect of scattering. It was determined that the scattering effect can not be detected because it is small compared to the effect of profile broadening.

\section*{Acknowledgments}
 We acknowledge insightful discussions with Sean Cutchin and
Roger Link. 
Construction of the LWA has been supported by the Office of Naval Research under Contract N00014-07-C-0147.  Support for operations and continuing development of the LWA1 is provided by the National Science Foundation under grants AST-1139963 and AST-1139974 of the University Radio Observatory program.  The computation is supported by the Advance Research Center of Virginia Tech.  Data reduction was performed using the BlueRidge system at Virginia Tech.

{\it Facility:} \facility{LWA}


\end{document}